\documentclass{webofc}

\usepackage[varg]{txfonts}   
\usepackage{hyperref}
\usepackage{url}
\hypersetup{colorlinks=true,citecolor=blue,urlcolor=blue,linkcolor=blue}
\usepackage{multicol}
\usepackage{multirow}
\usepackage{graphicx,subfigure}
\usepackage{upgreek}
\usepackage{lineno}     
\modulolinenumbers[2]
\switchlinenumbers   

\begin{document}
\title{Testing CPT symmetry via precision mass measurements of multi-strange baryons in ALICE}
%
%

\author{\firstname{Romain} \lastname{Schotter}\inst{1}\fnsep\thanks{\email{romain.schotter@cern.ch}} on behalf of the ALICE Collaboration. 
}

\institute{Stefan Meyer Institute for Subatomic Physics, Austrian Academy of Sciences.}

\abstract{
In these proceedings, the measurements of the $\Xi^{-}$, $\overline{\Xi}^{+}$, $\Omega^{-}$, $\overline{\Omega}^{+}$ masses and the mass differences between particle and anti-particle have been measured in pp collisions collected by the ALICE Collaboration during LHC Run 2. The results significantly improve the precision from previous experiments, thus allowing direct tests of CPT symmetry to an unprecedented level of precision in the multi-strange baryon sector.}
\maketitle
\section{Introduction}
\label{intro}
Fundamental symmetries stand as one of the most fruitful concepts in Modern Physics. They are of two kinds: continuous --- such as the global translations in both space and time, or the Lorentz transformations --- and discrete --- for example, the space- (P) and time- (T) inversions, the charge conjugation (C), and their combined transformation given by CPT. In particular, the Lorentz and CPT symmetries are closely connected by the so-called CPT theorem which states that any local Lorentz-invariant quantum field theory must also be CPT invariant \cite{kosteleckyStatusCPT1998}. Consequently, CPT violation implies the breaking of the Lorentz symmetry, and vice versa \cite{sozziTestsDiscreteSymmetries2019}. Another implication of CPT symmetry is that matter and antimatter share the same properties (invariant mass, lifetime, etc). Most of the experimental checks of CPT invariance stem from these physical consequences.

The Particle Data Group (PDG) \cite{particledatagroupReviewParticlePhysics2022} compiles a large variety of CPT tests from many experiments and with a high degree of precision; so far, no CPT violation has been observed. However, for a certain number of them, there is some room for improvement. Most notably, we can mention the measurements of the mass difference between particle and anti-particle in the multi-strange baryon sector. The only test of this nature dates back to 2006 \cite{abdallahMassesLifetimesProduction2006} for the $\Xi^{-}$ and $\overline{\Xi}^{+}$, and from 1998 \cite{chanMeasurementPropertiesOverline1998} for the $\Omega^{-}$ and $\overline{\Omega}^{+}$. As shown in Tab.~\ref{tab:PDGcharac}, both studies suffer from limited statistics. A similar observation can be made about the mass values, and particularly for the $\Omega$ baryons where the measurements rely on no more than 100 $\Omega$.

In these proceedings, we present a measurement of the mass difference of the $\Xi^{-}$ and $\overline{\Xi}^{+}$, and of the $\Omega^{-}$ and $\overline{\Omega}^{+}$ baryons. The data samples are much larger than those exploited previously: $\sim$2 400 000 ($\Xi^{-} + \overline{\Xi}^{+}$) and $\sim$130 000 $(\Omega^{-} + \overline{\Omega}^{+})$ with little background. These direct measurements of the mass difference offer a test of CPT invariance to an unprecedented precision in the multi-strange baryon sector. The absolute masses are updated as well, with a precision substantially better than the previous measurements. 

\begin{table}[h]
    \centering
	\caption{Particle properties with the last mass and mass difference measurements as of 2022, listed into~\cite{abdallahMassesLifetimesProduction2006, chanMeasurementPropertiesOverline1998, hartouniNclusiveRoductionEnsuremath1985, particledatagroupReviewParticlePhysics2022}. Here, the mass difference refers to the normalized one, namely $(M_{\overline{\text{part}}} - M_{\text{part}}) / M_{\text{average}}$.}\label{tab:PDGcharac}
    \footnotesize
    \begin{tabular}{cccc|cc}
    \noalign{\smallskip}\hline \hline \noalign{\smallskip}
    \multirow{3}{*}{Particle} & \multirow{3}*{Quark content} & \multirow{3}*{Mass measurement (MeV$/c^{2}$)} & \multirow{3}*{Sample} & Relative mass & \multirow{3}*{Sample}\\
    & & & & difference measurement & \\
    & & & & ($\times 10^{-5}$) & \\
    \noalign{\smallskip}\hline \noalign{\smallskip}
    $\Xi^{-}$ & $dss$ & 1321.70 $\pm$ 0.08 (stat.) $\pm$ 0.05 (syst.) & 2500 & \multirow{2}*{2.5 $\pm$ 8.7 (tot.)} & 2500\\
	$\overline{\Xi}^{+}$ & $\bar{d}\bar{s}\bar{s}$ & 1321.73 $\pm$ 0.08 (stat.) $\pm$ 0.05 (syst.) & 2300 & & 2300\\
    \noalign{\smallskip}\hline \noalign{\smallskip}
    $\Omega^{-}$ & $sss$ & 1672 $\pm$ 1 (tot.) & 100 & \multirow{2}*{1.44 $\pm$ 7.98 (tot.)} & 6323\\ 
    $\overline{\Omega}^{+}$ & $\bar{s}\bar{s}\bar{s}$ & 1673 $\pm$ 1 (tot.) & 72 & & 2607 \\ 
	\noalign{\smallskip}\hline \hline \noalign{\smallskip}
    \end{tabular}
\end{table}


\section{Detector setup and data sample}
\label{sec-1}

All the aforementioned particles are detected at mid-rapidity ($|y| < 0.5$), using the central detectors of ALICE \cite{alicecollaborationALICEExperimentCERN2008} at the LHC. The primary and secondary vertices are reconstructed using the Inner Tracking System (ITS), composed of six concentric layers of silicon detectors. The main tracking device is the Time Projection Chamber (TPC), which also provides particle identification of pions, kaons and protons based on their energy loss in the detector. The central part of the experiment is embedded in a large solenoid magnet (also called the L3 magnet), which offers three magnetic field configurations: +0.5, -0.5 and -0.2 T.

The mass of the particles of interest is measured in pp collisions at a centre-of-mass energy $\sqrt{s}$ = 13 TeV, using approximately $2.2 \times 10^9$ minimum-bias events collected in 2016, 2017 and 2018. Only data taken with a magnetic field value of $\pm$ 0.5 T are considered.

\section{Data analysis and study of the systematic uncertainties}
\label{sec-2}

In this measurement, the charged $\Xi$ and $\Omega$ baryons are studied in their cascade decay channel: $\Xi^{\pm} \rightarrow \uppi^{\pm} \Lambda \rightarrow \uppi^{\pm} \uppi^{\pm} {\rm p}^{\mp}$ (with a branching ratio BR = 63.9 \%) and $\Omega^{\pm} \rightarrow {\rm K}^{\pm} \Lambda \rightarrow {\rm K}^{\pm} \uppi^{\pm} {\rm p}^{\mp}$ (BR~=~43.4 \%) \cite{particledatagroupReviewParticlePhysics2022}. The reconstruction of these decay topologies is achieved by first reconstructing the $\Lambda$ candidates, which are then matched with a pion or kaon track. To reduce the induced combinatorial background, various topological and kinematic cuts are used, similar to the methods described in \cite{alicecollaborationMultiplicityDependenceMulti2020}.

The masses of the multi-strange baryons are measured through a fit of their invariant mass distributions, using the sum of two functions: a triple Gaussian for the peak and an exponential for the background. The measured mass corresponds to the center of the invariant mass peak, given by the position of the maximum of the triple Gaussian function, denoted as~$\mu$. The width ($\sigma$) provides an estimation of the mass resolution.

\begin{figure}[t]
\subfigure[]{
	\includegraphics[width=6.5cm,clip]{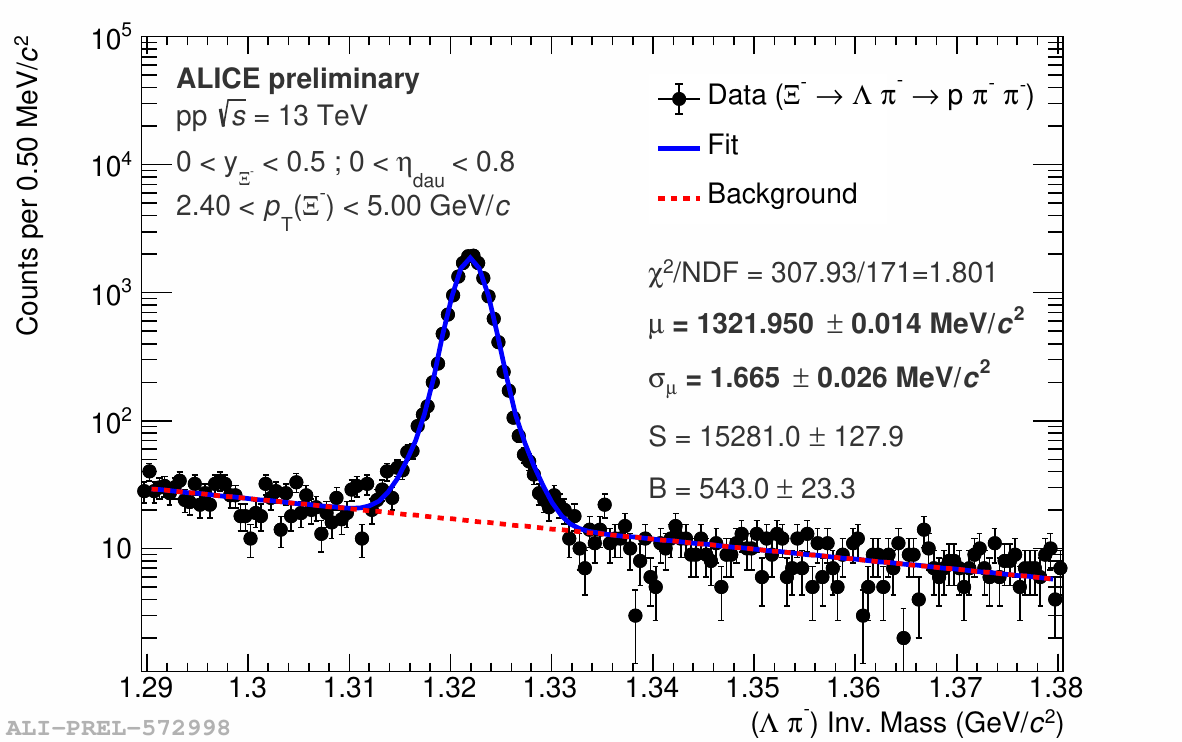}
	\label{fig:XiMinus_TripleGaussian}
} 
\subfigure[]{
	\includegraphics[width=6.5cm,clip]{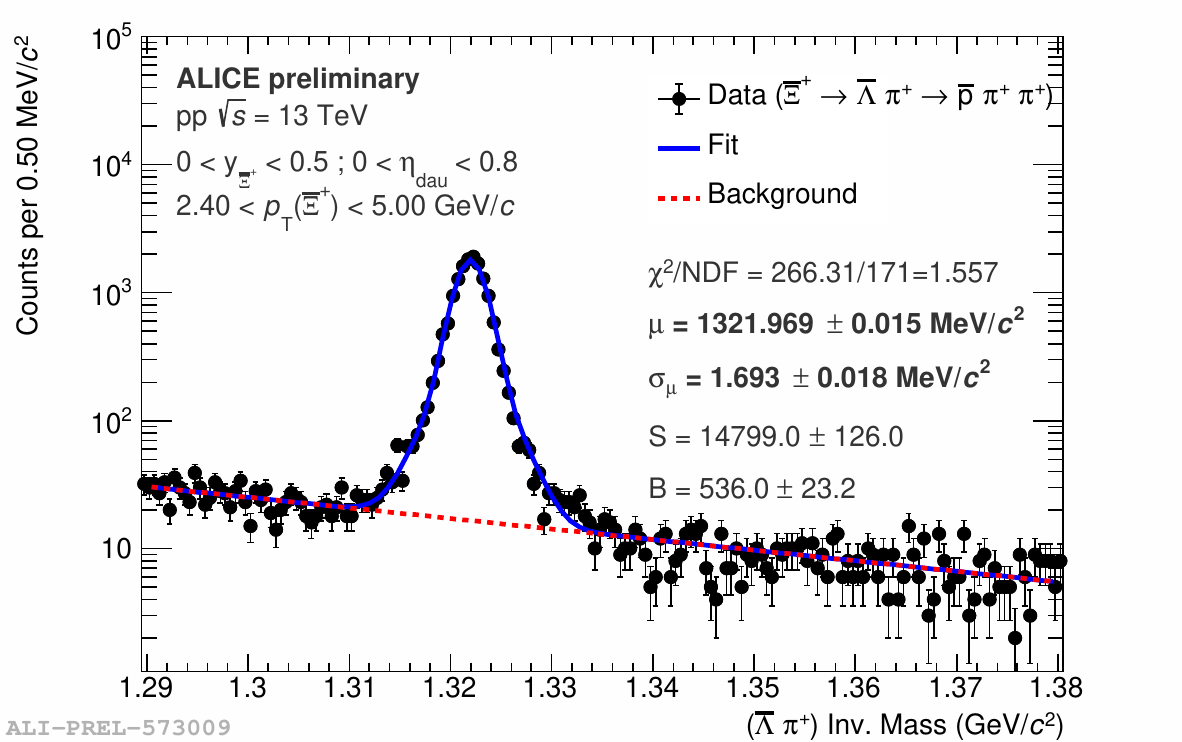}
	\label{fig:XiPlus_TripleGaussian}
} 
\subfigure[]{
	\includegraphics[width=6.5cm,clip]{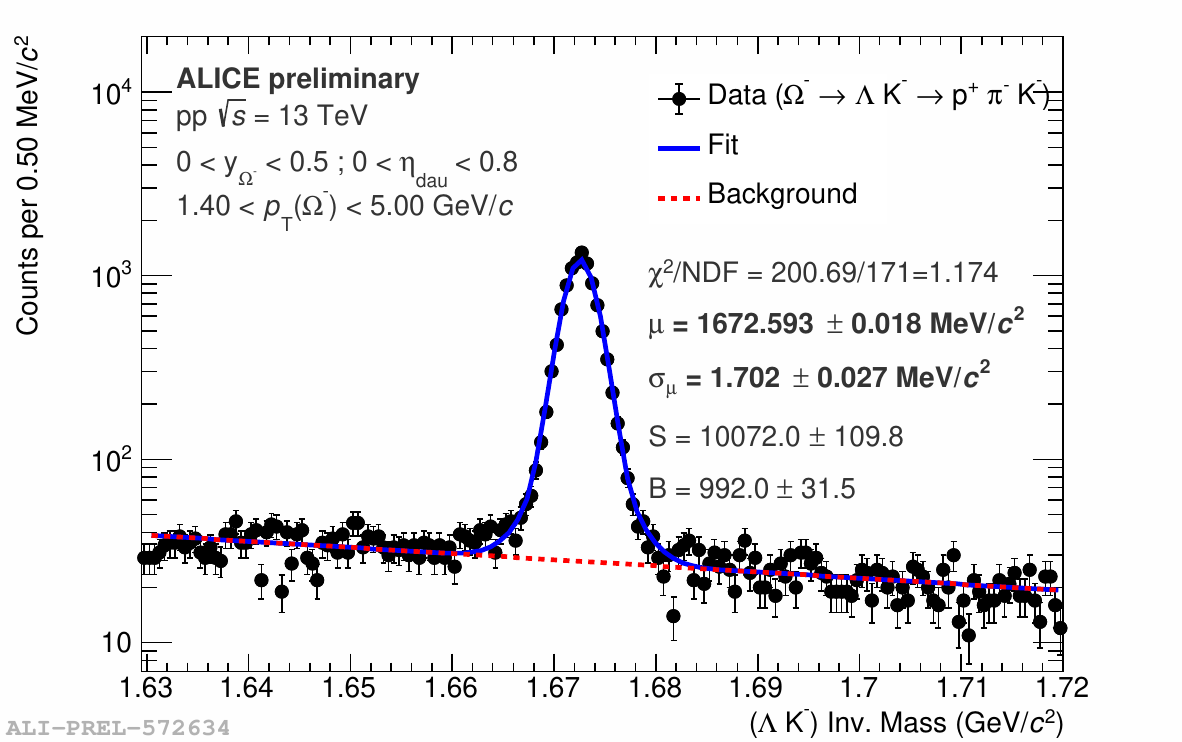}
	\label{fig:OmegaMinus_TripleGaussian}
} 
\subfigure[]{
	\includegraphics[width=6.5cm,clip]{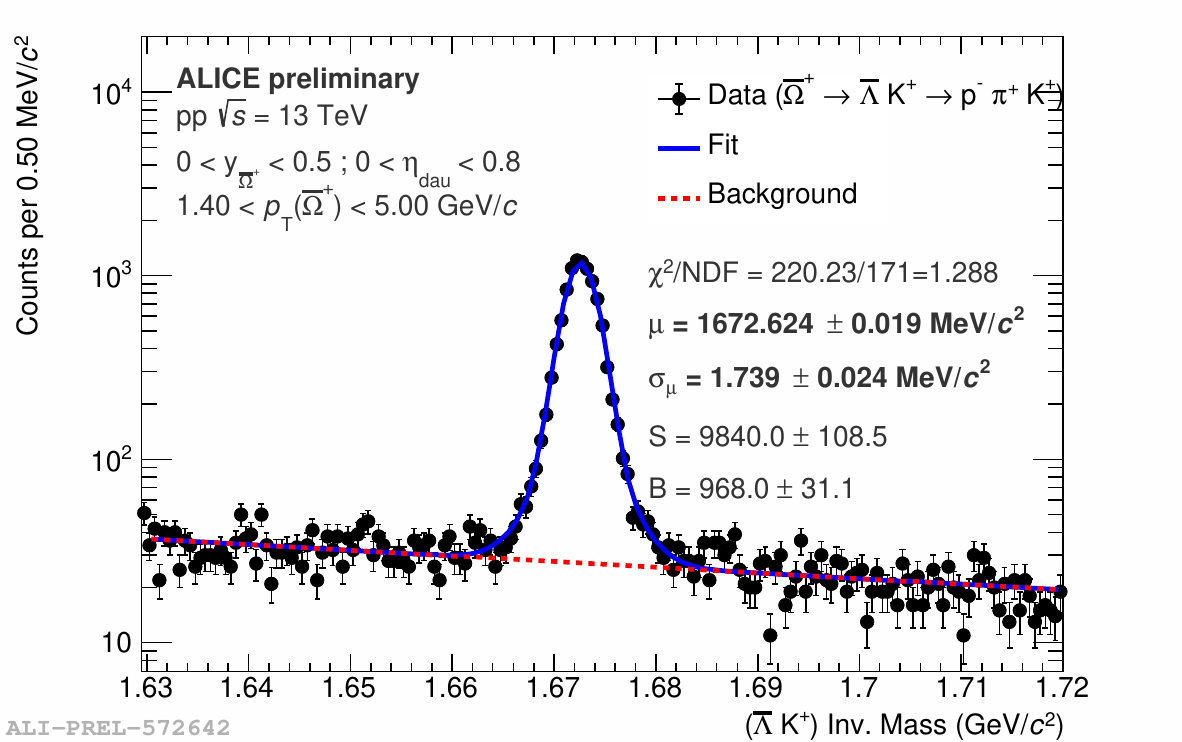}
	\label{fig:OmegaPlus_TripleGaussian}
} 
\caption{Invariant mass distributions of the $\Xi^{-}$ (\ref{fig:XiMinus_TripleGaussian}), $\overline{\Xi}^{+}$ (\ref{fig:XiPlus_TripleGaussian}), $\Omega^{-}$ (\ref{fig:OmegaMinus_TripleGaussian}) and $\overline{\Omega}^{+}$ (\ref{fig:OmegaPlus_TripleGaussian}). The peak is modelled by a triple Gaussian function, and the background by an exponential function. The measured mass and mass resolution, with their associated statistical uncertainties, are displayed in bold font.}\label{fig:InvMass}
\end{figure}

Figure~\ref{fig:InvMass} presents the invariant mass distributions of the $\Xi^{-}$, $\overline{\Xi}^{+}$, $\Omega^{-}$, $\overline{\Omega}^{+}$ in pp collisions at $\sqrt{s} = 13$ TeV. Considering the reduced $\chi^2$ values, all display a reasonably good fit. To ensure stable mass measurements, additional selections were applied. For instance, to deal with the residual distortions in the TPC, the analysis focuses solely on the positive $z$ side of the detector where they are less pronounced. Following these selections, the combinatorial background is greatly reduced, and overall 15 281 $\pm 128 \ \Xi^{-}$ (14 799 $\pm 126 \ \overline{\Xi}^{+}$) and 10 072 $\pm 110 \ \Omega^{-}$ (9 840 $\pm 109 \ \overline{\Omega}^{+}$) baryons were reconstructed. Although the final measurement relies only on a fraction of the initial data sample, the present results are still based on a statistics of strange baryons that is much larger than in previous measurements. 

The dominant sources of systematic uncertainties originate from the candidate selections, the detector calibration, the finite precision on the magnetic field map, and the limited knowledge on the material distribution. Other systematic effects have also been studied, namely the mass extraction procedure, the contribution of pile-up, the precision on the tabulated masses of the decay daughters, and the correction on the mass offset in simulation. However, their contributions remain small and do not exceed 20 keV$/c^{2}$.
\vspace*{-0.1cm}


\section{Results}

The final values of the $\Xi^{\pm}$ and $\Omega^{\pm}$ masses are:
\begin{align*}
    M(\Xi^{-}) &= 1321.975 \pm 0.026 \text{ (stat.)} \pm 0.078 \text{ (syst.)} \ \text{MeV/}c^2 ,\\
    M(\overline{\Xi}^{+}) &= 1321.964 \pm 0.024 \text{ (stat.)} \pm 0.083 \text{ (syst.)} \ \text{MeV/}c^2 ,\\
    M(\Omega^{-}) &= 1672.511 \pm 0.033 \text{ (stat.)} \pm 0.102 \text{ (syst.)} \ \text{MeV/}c^2 ,\\
    M(\overline{\Omega}^{+}) &= 1672.555 \pm 0.034 \text{ (stat.)} \pm 0.102 \text{ (syst.)} \ \text{MeV/}c^2 .
\end{align*}
The final relative mass difference between particle and anti-particle are:
\begin{align*}
    2 \times \frac{M(\overline{\Xi}^{+}) - M(\Xi^{-})}{M(\overline{\Xi}^{+}) + M(\Xi^{-})} &= \left[ -1.45 \pm 6.25 \text{ (tot.)} \right] \times 10^{-5} ,\\
    2 \times \frac{M(\overline{\Omega}^{+}) - M(\Omega^{-})}{M(\overline{\Omega}^{+}) + M(\Omega^{-})} &= \left[ 3.28 \pm 4.47 \text{ (tot.)} \right ] \times 10^{-5},
\end{align*}where the total uncertainty is given by the quadratic sum of the statistical and systematic ones.

\section{Discussion and conclusion}

Our measurements can be compared to the previously measured values presented in Tab.~\ref{tab:PDGcharac}. The uncertainty on the mass values has been reduced by approximately 15\% for the $\Xi$ and a factor 10 for the $\Omega$ baryons. On the mass difference, the precision has been improved by 40\% and almost a factor 2 for $\Xi$ and $\Omega$ baryons, respectively. 


Considering our precision, the relative mass difference measurements of $\Xi$ and $\Omega$ baryons are still compatible with 0 and further constrain the validity of CPT symmetry in the multi-strange baryon sector. While the $\Omega$ masses still agree with the PDG mass values in spite of the 10-fold improvement, the $\Xi$ masses are $2.5 \sigma$ larger than the tabulated values. 


These may influence the study of QCD, particularly in the non-perturbative regime. For instance, lattice QCD relies on the mass of multi-strange baryons to convert its predictions in lattice units into physical units. We can think about the $(g_{\mu} - 2)$ prediction of the BMW Collaboration, using the mass of the $\Omega$ baryons \cite{borsanyiLeadingHadronicContribution2021}, where our more precise mass values would lead all uncertainties from the physical input to be negligible; or the hadron mass spectroscopy that employs the mass of either $\Xi$ or $\Omega$ baryons \cite{fodorLightHadronMasses2012}, which still needs to be consistent with the measured spectrum using our masses to set the physical scale.

 \bibliography{Exported_Items} 
%
%

\end{document}